\documentstyle[12pt]{article}
\newcommand{\be}{\begin{eqnarray}}
\newcommand{\ee}{\end{eqnarray}}

\begin{document}

\rightline{}
\rightline{}
\rightline{}
\rightline{INFN-ISS 94/3}
\rightline{March 94}

\vspace{2cm}

\begin{center}

\LARGE{Hard constituent quarks and electroweak properties of pseudoscalar
mesons}\\

\vspace{2cm}

\large{ F. Cardarelli$^{(a)}$, I.L. Grach$^{(b)}$, I.M. Narodetskii$^{(b)}$\\
E. Pace$^{(a)}$, G. Salm\`{e}$^{(c)}$, S. Simula$^{(c)}$}\\

\vspace{1cm}

\normalsize{$^{(a)}$Department of Physics, University of Rome "Tor Vergata"\\
and INFN Sezione Tor Vergata\\
Via della Ricerca Scientifica, I-00133 Roma, Italy\\
$^{(b)}$Institute of Theoretical and Experimental Physics\\
Moscow 117259, Russia\\
$^{(c)}$Istituto Nazionale di Fisica Nucleare, Sezione Sanit\`{a},\\
Viale Regina Elena 299, I-00161 Roma, Italy}

\end{center}

\vspace{0.5cm}

\begin{abstract}
The high momentum components generated in the wave function of pseudoscalar
mesons
by the one-gluon-exchange interaction are investigated within a
relativistic constituent quark model. Adopting the light-cone formalism, the
sensitivity of the weak decay constant and the charge form factor to hard
constituent quarks is illustrated. \end{abstract}

\vspace{2cm}

Phys. Lett. B, in press.

\newpage
\pagestyle{plain}

\indent The investigation of the electroweak properties of mesons and baryons
can shed light on the effective constituents of hadrons and their dynamics. In
particular, the understanding of hadron electroweak properties has recently
received much theoretical attention within the context of constituent quark
models both in the framework of light-cone dynamics \cite{1,2,3,4} and in a
diagrammatic approach \cite{FM}. It has also been argued that existing
experimental data on the weak decay constant and the charge form factor of
mesons
can be accounted for within a constituent $q\bar{q}$ picture \cite{4}. However,
such a conclusion has been obtained by adopting a simplified description of the
dynamics of the constituent quarks inside the meson, namely, by using
gaussian-like wave functions which are expected to describe the effects of the
confinement scale only. In this letter the wave functions of pseudoscalar
(S-wave) mesons are analyzed within the relativized constituent quark model of
ref. \cite{5}, which properly describes meson (as well as baryon \cite{6}) mass
spectra in terms of an effective (QCD motivated) $q \bar{q}$ interaction
composed
by a one-gluon-exchange term and a linear confining potential. It is shown that
high momentum components, generated in the meson wave function by the
one-gluon-exchange term, sharply affect the weak decay constant and the charge
form factor. The agreement with existing experimental data can be obtained by
introducing an axial-vector coupling constant and a charge form factor of the
constituent quark. As in refs.~\cite{1,2,3,4}, meson electroweak properties
will
be evaluated adopting the light-cone formalism \cite{7}, which represents the
natural framework for constructing a relativistic quark model featuring the
dominance of the $q \bar{q}$ component of mesons \cite{FS}.

\indent {\bf 1. Light-cone meson wave functions.} As is known, light-cone wave
functions are eigenfunctions of the mass operator, e.g.
 \be
    M = M_0 + V
 \ee
and of the usual angular momentum operators $j^2$ and $j_n$, where the vector
$\hat{\bf n} = (0,0,1)$ defines the spin quantization axis. In eq. (1) $V$ is a
Poincar\'e invariant interaction term and $M_0$ is the free mass operator,
which
reads as
 \be
    M_0^2 = {k_{\perp}^2 + {m_q}^2 \over \xi} + {k_{\perp}^2 + {m_{\bar{q}}}^2
    \over 1 - \xi }
 \ee
where $m_q$($m_{\bar{q}}$) is the constituent quark (antiquark) mass and the
intrinsic light-cone variables $\vec{k}_{\perp}$ and $\xi$ are
 \be
    \xi= p_q^+/P^+ = 1 - p_{\bar{q}}^+/P^+
    \\
    \vec{k}_{\perp}=\vec{p}_{q \perp} - \xi \vec{P}_{\perp}=
    - \vec{p}_{\bar{q} \perp} + (1 - \xi)\vec{P}_{\perp}
 \ee
The subscript $\perp$ indicates the projection perpendicular to the spin
quantization axis and the {\em plus} component of a 4-vector $p \equiv (p^0,
{\bf p} )$ is given by $p^+ = p^0 + \hat{{\bf n}} \cdot {\bf p}$; in eqs. (3-4)
$\vec{P} \equiv (P^+, \vec{P}_{\perp}) = \vec{p}_q + \vec{p}_{\bar{q}}$ is the
total momentum of the meson, and $\vec{p}_q (\vec{p}_{\bar{q}})$ is the quark
(antiquark) momentum.  The structure of $M_0$ is more transparent if
the fraction $\xi$ is replaced by the longitudinal momentum $k_n$ defined as
 \be
     k_n = (\xi - {1 \over 2}) M_0 + {m_{\bar{q}}^2 - m_q^2 \over 2M_0}
 \ee
In terms of the new variable ${\bf k} \equiv (\vec{k}_{\perp}, k_n)$ the free
mass  operator is simply given by
 \be
    M_0 = \sqrt{m_q^2 + k^2 } + \sqrt{m_{\bar{q}}^2 + k^2 }
 \ee
with $k^2 = k_{\perp}^2 + k_n^2$.

In this letter pseudoscalar S-wave mesons will be considered. Omitting for sake
of simplicity the flavour and colour degrees of freedom, the intrinsic
light-cone wave function has the following structure (cf. refs. \cite{1,2,3,4})
 \be
    \Psi(\vec{k}_{\perp}, \xi,\nu \bar{\nu}) = R(\vec{k}_{\perp}, \xi, \nu
    \bar{\nu}) { w(k^2) \over \sqrt{4 \pi} } \sqrt{ {M_0 \over 4 \xi (1 - \xi)}
    \left [ 1 - \left ( m^2_q - m^2_{\bar{q}} \over M^2_0 \right ) ^2 \right ]
}
 \ee
where $\nu , \bar{\nu} = \pm 1/2$ are the spin projection variables and the
momentum-dependent spin factor $R$ is
 \be
    R(\vec{k}_{\perp}, \xi, \nu \bar{\nu}) = \sum_{\nu' \bar{\nu'}} ~ \langle
    \nu | R_M^{\dag} (\vec{k}_{\perp}, \xi, m_q) | \nu' \rangle ~ \langle
\bar{\nu}
    | R_M^{\dag} ( - \vec{k}_{\perp}, 1 - \xi, m_{\bar{q}}) | \bar{\nu'}
\rangle
    ~ \langle {1 \over 2} \nu' {1 \over 2} \bar{\nu'} | 00 \rangle
 \ee
In eq. (8) the $2\times 2$ irreducible representation of the Melosh rotation
\cite{8} reads as follows
 \be
    \langle \nu' | R_M (\vec{k}_{\perp}, \xi, m) | \nu \rangle =
\chi_{\nu'}^{\dag}
    { m + \xi M_0 - i \sigma \cdot (\hat{{\bf n}} \times {\bf k})
    \over \sqrt{(m + \xi M_0)^2 + k_{\perp}^2}} \chi_{\nu}
 \ee
where $\chi_{\nu}$ is the usual two-component Pauli spinor. The light-cone wave
function (7) satisfies the following normalization condition
 \be
    \int d \vec{k}_{\perp} d \xi ~ \sum_{\nu \bar{\nu}}
\Psi^{\dag}(\vec{k}_{\perp},
    \xi, \nu \bar{\nu}) \Psi(\vec{k}_{\perp}, \xi, \nu \bar{\nu}) =
\int_0^{\infty}
    dk k^2 | w(k^2) |^2 = 1
 \ee
Following the Brodsky-Huang-Lepage prescription \cite{9}, the radial wave
function
$w(k^2)$ appearing in eq.~(7) is identified with the equal-time wave function
in
the meson rest-frame. In this work we will adopt the effective $q \bar{q}$
Hamiltonian proposed by Godfrey and Isgur (GI) \cite{5} for the description of
meson mass spectra, viz.
 \be
    H_{q \bar{q}} ~ w(k^2) | 00 \rangle & \equiv & \left [ \sqrt{m_q^2 + k^2 }
+
    \sqrt{m_{\bar{q}}^2 + k^2 } + V_{q \bar{q}} \right ] ~ w(k^2) | 00 \rangle
    \nonumber \\
    & = & M_{q \bar{q}} w(k^2) | 00 \rangle
 \ee
where $M_{q \bar{q}}$ is the mass of the meson, $| 00 \rangle = \sum_{\nu
\bar{\nu}} \langle {1 \over 2} \nu {1 \over 2} \bar{\nu} | 00 \rangle
\chi_{\nu}
\chi_{\bar{\nu}}$ is the equal-time spin factor and $V_{q \bar{q}}$ is the
effective $q \bar{q}$  potential. The interaction in  the GI scheme,
$V_{(GI)}$,
is composed by a one-gluon-exchange term (dominant at short separations) and a
flavour-independent linear-confining term (dominant at large separations). In
order to analyze the effects of different terms of the GI $q \bar{q}$
interaction, two other choices of $w(k^2)$ will be considered; the first one is
the solution of eq.~(11) obtained after switching off the one-gluon-exchange
part
of  $V_{(GI)}$, i.e., by retaining only its confining part, $V_{(conf)}$,
whereas
the second choice is given by the solution of eq.~(11) in which only the
spin-independent part,$V_{(si)}$, of $V_{(GI)}$ is retained. The
three different forms of $w(k^2)$ will be denoted hereafter by $w_{(GI)}$,
$w_{(conf)}$ and $w_{(si)}$, corresponding to $V_{(GI)}$, $V_{(conf)}$ and
$V_{(si)}$, respectively. The wave equation (11) is solved by expanding the
wave
function $w(k^2)$ onto a (truncated) set of harmonic oscillator (HO) basis
states, viz.
 \be
    w(k^2) = \sum_{n=0}^{Q_{max}} b_n R_{n,0} (k^2)
 \ee
where $R_{n,\ell}$ is the usual HO radial function, $(n,\ell)$ are the HO
quantum
numbers and $Q_{max}$ is the maximum number of HO excitation quanta included in
the expansion (12). The coefficients $b_n$ are linear variational parameters,
which are determined by applying to the Hamiltonian $H_{q \bar{q}}$ the
Rayleigh-Ritz variational principle within the ansatz (12). We have checked
that
the value $Q_{max}=38$ ensures a complete convergence for all the quantities
considered in this work. As in ref. \cite{5}, the values $m_u = m_d = 0.220~
GeV$, $m_s = 0.419~ GeV$, $m_c = 1.628~ GeV$ and  $m_b = 4.977~ GeV$ are
adopted. We have also checked that, using the full GI $V_{q \bar{q}}$
interaction,
the calculated mass spectra of pseudoscalar mesons coincide with the results of
ref. \cite{5}. The three wave functions $w_{(GI)}$, $w_{(conf)}$ and $w_{(si)}$
for the pion are compared in fig. 1: it can be seen that $w_{(GI)}$ is
characterized by high momentum components ($k > 0.5 ~ GeV/c$) which are totally
absent in $w_{(conf)}$, whereas the effects of the spin-spin part of the
one-gluon-exchange interaction strongly modifies the wave function at $k > 1
{}~GeV/c$. It should be pointed out that  $w_{(conf)}$ is similar to the
gaussian-like wave function adopted in refs. \cite{1,2,3,4}. The wave functions
$w_{(GI)}$ of various pseudoscalar mesons are reported in fig. 2, where it can
be
seen that the high momentum tail of the wave function is similar in light and
heavy pseudoscalar mesons. To sum up, the  short-range structure of the
effective
$q \bar{q}$ interaction, determining the hyperfine splitting of hadron mass
spectra, sharply affects the high momentum components of meson wave functions,
i.e. it originates the presence of hard constituent quarks in mesons.

 \indent {\bf 2. Meson electroweak properties.} Let us now consider the charge
form factor $F^{PS}(Q^2)$ and the axial-vector weak decay constant $f_A^{PS}$
of
pseudoscalar mesons. As for the latter quantity, following the light-cone
approach of ref. \cite{4} and introducing an axial-vector coupling constant at
the level of constituent quarks, $g_A^q(0)$ [2b], one
gets
 \be
    f_A^{PS} & = & g_A^q(0) {\sqrt{6} \over (2\pi)^{3/2}}~ \int ~
    d\vec{k}_{\perp}~ d\xi~ \sqrt{ {M_0 \over 4 \xi (1 - \xi)} \left [ 1 -
\left
    ( m^2_q - m^2_{\bar{q}} \over M^2_0 \right ) ^2 \right ] }
    \nonumber \\
    & ~ & {w(k^2) \over \sqrt{4\pi}}~ {(1 - \xi)~m_q + \xi~m_{\bar{q}} \over
    \sqrt{\xi(1 - \xi)}~ \sqrt{M^2_0 - (m_q-m_{\bar{q}})^2}}
 \ee
The results of the calculations of $f_A^{PS}$ performed using $w_{(GI)}$
and $w_{(conf)}$ in eq.~(13), are reported in table 1 and compared with
existing
experimental data \cite{10,11} and lattice QCD calculations \cite{12,13,14}. It
can be seen that: i) $f_A^{PS}$ increases in  presence of hard constituent
quarks; ii) if $g_A^q(0)=1$ is used (as for the current quarks), the results
obtained by considering only soft constituent quarks (i.e., by retaining only
the
confining term of the GI $q \bar{q}$ interaction) account for the data on $\pi$
and $K$ mesons, but they do not reproduce the lattice QCD predictions for
heavier
mesons; iii) in presence of hard constituent quarks (i.e., when the full GI $q
\bar{q}$ interaction is taken into account), a reasonable agreement both with
the
data and the lattice QCD calculations can be obtained for all the considered
pseudoscalar mesons  by choosing $g_A^q(0) \neq 1$ and flavour independent
(note
that in table 1 the value $g_A^q(0) = 0.726$ has been chosen in order to
reproduce the central value of the experimental data for $f_A^{\pi}$). It
should
be pointed out that our predictions for the ratio  $f_A^{PS} / f_A^{\pi}$,
obtained using the full GI $q \bar{q}$ potential, compare favourably with the
data
and the lattice QCD calculations; this result depends only upon the GI wave
function of the mesons, if $g_A^q(0)$ is assumed to be flavour independent.
Furthermore, our results for the two ratios $f_A^{D_s} / f_A^D$ and $f_A^{B_s}
/
f_A^B$ are $1.17$ and $1.20$, respectively,  in good agreement with the
corresponding predictions $1.21 \pm 0.06$ and $1.22 \pm 0.02$ obtained from QCD
sum rules \cite{DP}.

Let us now investigate the sensitivity of the charge form factor of $\pi$
and $K$ mesons to soft and hard constituent quarks. As is known \cite{1,2,3,4},
within the light-cone formalism the charge form factor is related to the matrix
elements of the so-called {\em good} component of the electromagnetic current
density, which is specified by its {\em plus} component $J^+$ when the spin
quantization axis is chosen so that the {\em plus} component of the
four-momentum
transfer vanishes ($Q^+ = 0$). Following refs. \cite{1,2,3,4} and introducing a
charge form factor of the constituent quark, $F^q(Q^2)$ [2b], one gets
 \be
    F^{PS}(Q^2) = e_q F^q(Q^2) H(Q^2, m_q, m_{\bar{q}} ) + e_{\bar{q}}
    F^{\bar{q}}(Q^2) H(Q^2, m_{\bar{q}}, m_q)
 \ee
where $e_q$ is the charge of the constituent quark and
 \be
    H(Q^2, m_1, m_2) & = & \int d\vec{k}_{\perp}~ d\xi ~ { \sqrt{M_0 M'_0}
\over
    4 \xi (1 - \xi) } ~ \sqrt{ \left [ 1 - \left( {m_1^2 - m_2^2 \over M_0^2 }
    \right ) ^2 \right ] ~ \left [ 1 - \left ( {m_1^2 - m_2^2 \over {M'}_0^2}
    \right ) ^2 \right ] }
    \nonumber \\
    & ~ & {w(k^2) ~ w^{\ast}({k'}^2) \over 4 \pi }~ {\cal{M}} (\vec{k}_{\perp},
    \vec{k'}_{\perp}, \xi, m_1, m_2)
 \ee
with
 \be
    \vec{k'}_\perp = \vec{k}_\perp + (1 - \xi) \vec{Q}_\perp \nonumber \\
    {M'}_0^2 = {{k'}_{\perp}^2 + m_1^2 \over \xi} + {{k'}_{\perp}^2 + m_2^2
\over
    (1 - \xi)} \nonumber \\
    k'_n = (\xi - {1 \over 2}) M'_0 + {m_2^2 - m_1^2 \over 2 M'_0} \nonumber
 \ee
 In eq. (15) ${\cal M}$ is the contribution of the Melosh rotations, which
reads
as
 \be
   {\cal{M}} (\vec{k}_{\perp}, \vec{k'}_{\perp}, \xi, m_1, m_2) = { \xi (1 -
   \xi)~ [M^2_0 - (m_1 - m_2)^2 ] + \vec{k}_\perp \cdot (\vec{k'}_\perp -
   \vec{k}_\perp) \over \xi (1 - \xi)~ \sqrt{M_0^2 - (m_1 - m_2)^2}~
   \sqrt{ {M'}_0^2 - (m_1 - m_2)^2} }
 \ee
The results of the calculations of $F^{\pi}$ (eq. (14)), performed using the
wave
functions $w_{(GI)}$, $w_{(conf)}$ and $w_{(si)}$ and $F^q=1$ (as for the
current quarks), are shown in figs. 3 and 4 and compared with existing
experimental data \cite{15,16,19} at low and high values of $Q^2$.  It can be
seen that:  i) the pion form factor is strongly affected by hard constituent
quarks both at low and high $Q^2$;  ii) the theoretical predictions based on
soft
constituent quarks only ($w_{(conf)}$), compare favourably with the
experimental
data for values of $Q^2$ up to several $(GeV/c)^2$, in agreement with the
results
of refs. \cite{1,2,3,4}. From fig. 3 it can also be seen that the  charge
radius
$<r^2>_{\pi} \equiv -  6 (dF^{\pi} / dQ^2)_{Q^2=0}$ of the pion is strongly
underestimated using the full GI $q \bar{q}$ potential. The agreement with the
experimental value (~$<r^2>_{\pi}^{exp} = (0.660 \pm 0.024~fm)^2$ ~[20b]) can
be
recovered by considering $F^q \neq 1$ in eq. (14). Let us assume a simple
monopole behaviour of the constituent quark form factor, viz.
 \be
    F^q(Q^2) = {1 \over 1 + Q^2 <r^2>_q / 6 }
 \ee
and fix $<r^2>_u = <r^2>_d = (0.48~fm)^2$ in order to reproduce the
experimental
value of the pion charge radius. The pion form factor obtained using in
eq.~(14)
$F^u = F^d$ given by eq.~(17) with $<r^2>_q = (0.48~fm)^2$, is represented by
the
solid line in figs. 3 and 4. It can clearly be seen that the introduction of a
single phenomenological form factor of the constituent quark is able to bring
theoretical predictions, based on the full GI $q \bar{q}$ interaction, in
agreement with data for values of $Q^2$ up to several $(GeV/c)^2$.  The results
for the kaon charge form factor are reported in fig. 5 and compared with
existing
experimental data \cite{21,22}. As in the case of the pion, we have considered
both $F^q = 1$ and  $F^q \neq 1$; in the latter case a flavour independent form
factor of the constituent quark, i.e. $F^u = F^d = F^s$, has been adopted using
the monopole form of eq.~(17) with $<r^2>_q = (0.48~fm)^2$ (i.e., the same
charge
radius of the constituent quark used for the pion). When $F^q \neq 1$ one gets
$<r^2>_{K^+} = 0.41~fm^2$ and $<r^2>_{K^0} = -0.036~fm^2$ to be compared with
the
experimental values $<r^2>_{K^+}^{exp} = 0.34 \pm 0.05~fm^2$ and
$<r^2>_{K^0}^{exp} = -0.054 \pm 0.026~ fm^2$ \cite{21}. We would like to stress
that the charge radius of the neutral kaon is independent of the charge radius
of
the constituent quark, if the latter is assumed to be flavour independent;
thus,
the agreement of our result of $<r^2>_{K^0}$ with the data depends only upon
the
GI wave function of the kaon. The slight disagreement among the experimental
and
the calculated values of $<r^2>_{K^+}$ might be ascribed to small deviations of
$F^s$ from the $SU(3)$ symmetric value $F^u=F^d=F^s$.

\indent In conclusion we have analyzed the wave function of pseudoscalar mesons
within the relativized constituent quark model of ref. \cite{5}. The large
amount
of high momentum components of the wave function, related to the
one-gluon-exchange part of the effective $q \bar{q}$ potential, sharply affects
the calculations of physical quantities, like the weak decay constant and the
charge form factor, performed within the framework of the light-cone
formalism. In particular, the ratio of the weak decay constant of flavoured
pseudoscalar mesons to that of the pion is in agreement with the existing
experimental data and the predictions of the fundamental theory; this fact
gives
us confidence in the overall behaviour of the Godfrey-Isgur wave functions. The
extension of our calculations to the nucleon electroweak properties is in
progress and should allow to check the validity of the introduction of an
axial-vector weak coupling constant and a charge form factor at the level of
the
constituent quark. As for the meson sector, the comparison of theoretical
predictions with precise experimental data in a wide range of $Q^2$ for charge
as
well as transition form factors (as planned at CEBAF) should provide unique
information on the $q \bar{q}$ component of meson wave functions.

\vspace{1cm}

We gratefully acknowledge T. Frederico and M. Strikman for enlightening
discussions.

\newpage
\vspace{2cm}
\begin{center}
{\bf Table Caption}
\end{center}
\vspace{2cm}
Table 1. Axial-vector weak decay constant $f_A^{PS}$ (eq. (13)) calculated
for various pseudoscalar mesons adopting different wave functions $w(k^2)$ (see
text) and different values of the axial-vector weak coupling constant
$g_A^q(0)$
for the constituent quarks. The results of the calculations are compared with
existing experimental data \cite{10,11} and lattice QCD calculations
\cite{12,13,14}.

\newpage
\vspace{2cm}
\begin{center}
{\bf Figure Captions}
\end{center}

\vspace{2cm}

Fig. 1. The function $| k \cdot w(k^2) |^2$ for the pion, plotted versus the
internal momentum $k$. Dot-dashed line: gaussian-like wave function of ref.
\cite{4}. Dotted line: wave function $w_{(conf)}$ obtained from eq.~(11) using
only the confinement part of the $q \bar{q}$ interaction of ref. \cite{5}.
Dashed
line: wave function $w_{(si)}$ solution of eq.~(11) neglecting the spin-spin
part
of the $q \bar{q}$ interaction of ref. \cite{5}. Solid line: wave function
$w_{(~GI)}$ obtained from eq.~(11)  using the full $q \bar{q}$ interaction of
ref. \cite{5}.

\vspace{1cm}

Fig. 2. The function $| k \cdot w(k^2) |^2$ for various pseudoscalar mesons,
obtained using in eq.~(11) the full $q \bar{q}$ interaction of ref. \cite{5}.
Dot-dashed: $\pi$; dotted line: $K$; dashed line: $D$; dashed line with full
dots: $D_S$; solid line: $B$; solid line with triangles: $B_S$.

\vspace{1cm}

Fig. 3. The square of the charge form factor of the pion (eq.~(14)) at low
values
of $Q^2$.  The dotted, dashed and dot-dashed lines represent the results of the
calculations performed using the wave functions $w_{(conf)}$, $w_{(si)}$ and
$w_{(GI)}$, respectively, and adopting $F^q=1$ (see text). The solid line is
the
result of the calculations of eq.~(14) obtained using $w_{(GI)}$ and the
monopole
form for  $F^q$ (eq.~(17) with $<r^2>_q = (0.48~fm)^2$). Experimental data are
from refs. \cite{15} (full dots) and \cite{19} (open dots).

\vspace{1cm}

Fig. 4. The charge form factor of the pion (eq.~(14)) versus $Q^2$. The dotted,
dashed and dot-dashed lines represent the results of the calculations
performed using the wave functions $w_{(conf)}$, $w_{(si)}$ and $w_{(GI)}$ ,
respectively, and adopting $F^q=1$ (see text). The solid line is the  result of
the calculations of eq.~(14) obtained using $w_{(GI)}$ and the monopole form
for
$F^q$ (eq.~(17) with $<r^2>_q = (0.48~fm)^2$). Experimental data are from refs.
\cite{15} (open dots), [19a] (full dots), [19b] (open squares) and [19c] (full
squares).

\vspace{1cm}

Fig. 5. The same as in fig. 3, but for the kaon charge form factor.
Experimental
data are from refs. \cite{21} (open dots) and \cite{22} (full dots).

\newpage
\begin{center}
\vspace{2cm}
{\bf TABLE 1}\\
\vspace{0.5cm}
\begin{tabular} {||c ||c |c |c ||c ||c ||}
\cline{1-6} \cline{1-6}
      ~&$conf$&    $GI$&     $GI$&    ~&   ~
\\ \cline{2-4}
 $meson$&  $g_A^q=1$&  $g_A^q=1$&  $g_A^q=0.726$&  $lattice~QCD$&  $exp.~data$
\\ \cline{1-6} \cline{1-6}
  $\pi$&  91.2&   127.3&    92.4&          ~&   92.4$\pm$0.2~~{\cite{10}}
\\ \cline{1-6} \cline{1-6}
  $K$&   108.6&   159.9&   116.1&    107$\pm$7~~{\cite{14}}&
                                                  113.4$\pm$1.1~~{\cite{10}}
\\ \cline{1-6} \cline{1-6}
   ~&        ~&       ~&       ~&    156$\pm$9~~{\cite{12}}&               ~
\\
 $D$&    124.4&   215.3&   156.3&    147$\pm$8~~{\cite{13}}&
                                                         $<$219~~{\cite{11}}
\\
   ~&        ~&       ~&       ~&    154$\pm$7~~{\cite{14}}&               ~
\\ \cline{1-6} \cline{1-6}
$D_s$&   142.1&   252.8&   183.5&    163$\pm$7~~{\cite{13}}&               ~
\\
    ~&       ~&       ~&       ~&    170$\pm$7~~{\cite{14}}&               ~
\\ \cline{1-6} \cline{1-6}
  $B$&   101.3&   184.1&   133.6&   141$\pm$16~~{\cite{12}}&               ~
\\
   ~&        ~&       ~&       ~&    132$\pm$9~~{\cite{13}}&               ~
\\ \hline
$B_s$&   116.8&   220.6&   160.1&    146$\pm$7~~{\cite{13}}&               ~
\\ \cline{1-6} \cline{1-6}
\end{tabular}
\end{center}
\end{document}